\documentclass[sigconf]{acmart}

\usepackage[T1]{fontenc}
\usepackage[latin1]{inputenc}

\usepackage{url}
\usepackage{amsmath}
\usepackage{xcolor} 
\usepackage{algorithm2e} 

\DeclareMathOperator*{\argmax}{arg\,max}

\usepackage{xcolor}
\definecolor{c_red}{HTML}{BE503E}
\definecolor{c_green}{HTML}{7BB292}
\definecolor{c_yellow}{HTML}{D7AE38}
\definecolor{c_blue}{HTML}{628B97}
\definecolor{c_purple}{HTML}{B48EAD}

\newcommand{\Ni}{(1)~}
\newcommand{\Nii}{(2)~}
\newcommand{\Niii}{(3)~}
\newcommand{\Niv}{(4)~}
\newcommand{\Nv}{(5)~}

\usepackage{booktabs}

\usepackage{graphicx}
\DeclareGraphicsExtensions{.pdf,.ai,.jpg,.png}
\setkeys{Gin}{pagebox=artbox}
\graphicspath{{./ictir22-pairwise-ranking-figures/}}

\newif\ifbscomment
\bscommentfalse
\bscommenttrue

\RequirePackage{xcolor}
\RequirePackage{soul}
\setstcolor{blue}

\newsavebox\bscombox
\newcommand{\bscom}[3][]{%
  \sbox{\bscombox}{\fontsize{8}{9}\selectfont#1#2#3}
  \noindent
  \st{#2}{\selectfont
    \color{blue}#3\ifx\\#1\\\else{\color{violet}[#1]}\fi
    }
  }

\begin{document}
\copyrightyear{2022} 
\acmYear{2022} 
\setcopyright{acmlicensed}\acmConference[ICTIR '22]{Proceedings of the 2022 ACM SIGIR International Conference on the Theory of Information Retrieval}{July 11--12, 2022}{Madrid, Spain}
\acmBooktitle{Proceedings of the 2022 ACM SIGIR International Conference on the Theory of Information Retrieval (ICTIR '22), July 11--12, 2022, Madrid, Spain}
\acmPrice{15.00}
\acmDOI{10.1145/3539813.3545140}
\acmISBN{978-1-4503-9412-3/22/07}

\title{Sparse Pairwise Re-ranking with Pre-trained Transformers}

\author{Lukas Gienapp}
\orcid{1234-5678-9012}
\affiliation{%
  \institution{Leipzig University}
  \city{}
  \country{}
}
\author{Maik Fr{\"o}be}
\orcid{1234-5678-9012}
\affiliation{%
  \institution{Martin-Luther-Universit{\"a}t Halle-Wittenberg}
  \city{}
  \country{}
}
\author{Matthias Hagen}
\orcid{1234-5678-9012}
\affiliation{%
  \institution{Martin-Luther-Universit{\"a}t Halle-Wittenberg}
  \city{}
  \country{}
}
\author{Martin Potthast}
\orcid{1234-5678-9012}
\affiliation{%
  \institution{Leipzig University}
  \city{}
  \country{}
}

\begin{abstract}
Pairwise re-ranking models predict which of two documents is more relevant to a query and then aggregate a final ranking from such preferences. This is often more effective than pointwise re-ranking models that directly predict a relevance value for each document. However, the high inference overhead of pairwise models limits their practical application: usually, for a set of $k$~documents to be re-ranked, preferences for all $k^2-k$~comparison pairs excluding self-comparisons are aggregated. We investigate whether the efficiency of pairwise re-ranking can be improved by sampling from all pairs. In an exploratory study, we evaluate three sampling methods and five preference aggregation methods. The best combination allows for an order of magnitude fewer comparisons at an acceptable loss of retrieval effectiveness, while competitive effectiveness is already achieved with about one third of the comparisons.
\end{abstract}

\ccsdesc[500]{Information systems~Learning to rank}
\ccsdesc[500]{Information systems~Rank aggregation}
\ccsdesc[500]{Information systems~Retrieval effectiveness}
\ccsdesc[500]{Information systems~Retrieval efficiency}

\keywords{Pairwise re-ranking; Sampling; Efficiency; Pre-trained transformers}

\maketitle

\section{Introduction}

\begin{figure}[t]
\centering
\includegraphics[width=\columnwidth]{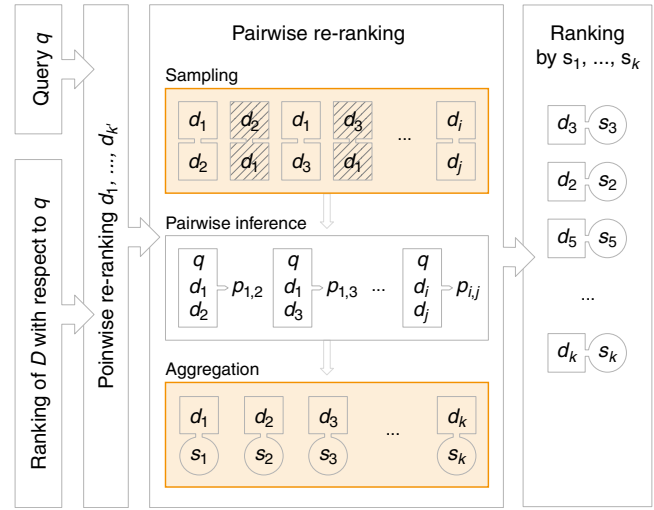}
\Description[Illustration of the mono-duo-design pattern for re-ranking.]{Illustration of the mono-duo-design pattern for re-ranking, where the pairwise re-ranking step is subdivided into the three steps sampling, pairwise inference, and aggregation}
\caption{The mono-duo design pattern for re-ranking. Parts investigated in this paper are highlighted in orange. Comparisons omitted by sampling are striped.}
\label{mono-duo-reranking-illustration}
\end{figure}

Pre-trained transformers have ushered in a new era in information retrieval: with a sufficient amount of training data, transformer-based re-ranking models can significantly outperform traditional retrieval models~\cite{lin:2019}. Two classes of re-rankers are implemented using pre-trained transformers~\cite{lin:2021}:
\Ni
pointwise re-rankers that predict the relevance of a document~$d$ to a query~$q$, and
\Nii
pairwise re-rankers that predict which of two documents~$(d_i, d_j)$ is more relevant to~$q$.
To further maximize re-ranking effectiveness, the mono-duo design pattern~\cite{pradeep:2021} shown in Figure~\ref{mono-duo-reranking-illustration} applies both sequentially. Given a query~$q$, a document set~$D$, and a ranking of~$D$ produced by a traditional retrieval model like~BM25, the top-$k'$ documents~$D_{k'}=\{d_1,\ldots,d_{k'}\}\subset D$ are re-ranked according to their pointwise relevance to~$q$. Then, the top-$k$ documents~$D_k\subset D_{k'}$,~$k\ll k'$, are re-ranked based on pairwise comparisons in three steps. First, pairs of documents~$(d_i, d_j)$ are sampled, where~$i,j \in [1,k]$ and~$i\neq j$. Second, each pair~$(d_i, d_j)$ is passed to a transformer model to predict a probability~$p_{ij}$ indicating whether the document~$d_i$~($p_{ij}\geq 0.5$) or the document $d_j$~($p_{ij}< 0.5$) is more relevant to~$q$. Third, for each document~$d_i$, a relevance score~$s_i$ is aggregated from all probabilities of comparisons including~$d_i$, and these scores are then used to derive the final pairwise re-ranking.

Empirical evidence suggests that pairwise re-rankers are more effective than pointwise re-rankers since their relevance scores take the relative relevance differences between documents into account, rather than making independent relevance predictions~\cite{pradeep:2021}. To maximize the potential effectiveness gains, previous work has relied on exhaustive comparisons of all $k^2-k$~pairs of the top-$k$ documents~$D_k$ to be re-ranked. Given the high run time overhead of transformer inferences, this quadratic step led to the recommendation that the re-ranking depth should be limited to~$k\leq 50$.

However, many of the estimated comparison probabilities may be redundant in that they can be predicted from those of other comparisons. A theoretical lower bound on the run time complexity is~$O(k\log k)$ using a suitable sorting algorithm if the estimated comparisons were ``consistent'' (i.e.,~$p_{ij} = 1 - p_{ji}$) and transitive. We investigate for the first time if the efficiency of pairwise re-rankers can be increased without a significant loss of effectiveness by sampling from and thus sparsifying the comparison set.

The two components of the mono-duo re-ranking pipeline that we study in this paper are highlighted in Figure~\ref{mono-duo-reranking-illustration}: We introduce a sampling step before the pairwise inference to draw a subset of the $k^2-k$~possible comparisons, and we revisit the aggregation step since its effectiveness directly depends on the kind of sample it receives (Section~\ref{sec:methodological-approach}). To investigate the effect of sparsification on the retrieval effectiveness, we study three sampling methods (global random, exhaustive window, skip-window) and five aggregation methods (sorting, summation, regression, greedy, and graph-based) on three datasets~(ClueWeb09, ClueWeb12, MS~MARCO) using the pointwise monoT5 and the pairwise duoT5 models~\cite{pradeep:2021}. Our results show that skip-window sampling with greedy aggregation allows for an order of magnitude fewer comparisons at an acceptable loss of effectiveness, while competitive effectiveness to the all-pairs approach can already be achieved with only one third of the comparisons~(Section~\ref{sec:results}). All code and data underlying our experiments are publicly available.%
\footnote{\url{https://github.com/webis-de/ICTIR-22}}

\section{Related Work}
\label{sec:related-work}

We briefly give some background on the history of learning-to-rank retrieval models before detailing the nature of pairwise learning-to-rank models and reviewing rank aggregation approaches, which we employ to aggregate pairwise preferences into a final ranking. Finally, we describe some related efforts at making transformer-based learning to rank more efficient.

\paragraph{Learning to Rank.}
Since decades, machine learning has been applied to improve retrieval effectiveness~\cite{fuhr:1989,lin:2021}. Traditional feature-based learning-to-rank models evolved from pointwise over pairwise to listwise approaches~\cite{liu:2011}. While feature-based models are still successful~\cite{qin:2021}, the recent promising retrieval effectiveness results of pre-trained transformer models~\cite{lin:2021} has shifted the community's focus away from feature-based learning to rank. But history appears to repeat itself as the aforementioned evolution from pointwise approaches like monoBERT~\cite{nogueira:2019} and monoT5~\cite{nogueira:2020} to pairwise approaches~\cite{lin:2021} can been observed as well.

\enlargethispage{\baselineskip}
\paragraph{Pairwise Learning to Rank.}
Pairwise learning-to-rank approaches predict which document in a pair is probably more relevant to a query and should be ranked higher~\cite{liu:2011}. In feature-based as well as transformer-based learning to rank, pairwise approaches usually outperform pointwise ones that score documents independently of each other~\cite{liu:2011}. Yet, when the inference step of a pairwise model compares all pairs of documents, the run time requirement is quadratic in the number of documents to be ranked. In an effort to reduce the comparison count, extensive studies of theoretical properties of feature-based pairwise approaches~\cite{clemencon:2008,lan:2012} have led to suggestions for better run time characteristics. For example, SortNet~\cite{rigutini:2011} uses a learned preference function that is guaranteed to output symmetric preferences, allowing to skip half of the comparisons. Yet, recent pairwise transformer-based models like duoBERT~\cite{pradeep:2021} and the more effective duoT5~\cite{pradeep:2021} lack the desirable symmetry property of models like SortNet. Additionally, the theoretical analysis of duoBERT and duoT5 is still in its infancy; previous work even found such models difficult to be interpreted~\cite{macavaney:2020,voelske:2021}. The effectiveness of duoBERT or duoT5 relies on computing preferences for {\em all} pairs of documents, at the expense of their efficiency~\cite{zhang:2021} limiting their applicability in search scenarios with run time constraints.

\paragraph{Rank Aggregation.}
Rank aggregation~\cite{liu:2011} uses pairwise relevance preference probabilities to derive a ranking---often by computing a score for each individual document. Finding an optimal aggregation with arbitrarily-sized inputs is an $\mathsf{NP}$-hard problem~\cite{cohen:1999}, but many approaches are known to work well in practice. While dynamic aggregation methods decide which documents to compare next based on all previous comparisons, static aggregation methods assume that the required pairwise comparisons are conducted before the actual aggregation starts~\cite{wu:2021}.

In our study, we employ the following five aggregation methods (details in Section~\ref{method:aggregation}): 
\Ni
Sorting via the KwikSort method~\cite{ailon:2008}, which assumes that the comparisons are consistent and form a total order,
\Nii
additive aggregation~\cite{pradeep:2021}, where the rank of a document is indicated by the sum of the document's comparison probabilities (potentially transforming the probabilities before summation),
\Niii 
regression-based aggregation~\cite{thurstone:1927,bradley:1952,stern:1992,zhou:2008}, where latent scores for documents are learned so that they optimally correspond to a given set of pairwise comparisons,
\Niv
greedy aggregation~\cite{cohen:1999, aledo:2021}, in which a heuristic iteratively selects and removes the best document from a given set and then proceeds with the rest, and
\Nv
graph-based aggregation~\cite{xiao:2021}, where comparisons are interpreted as directed edges between document nodes and a measure of graph centrality is used to derive a ranking score. 

\paragraph{Efficiency Improvements for Transformer-based Re-Rankers.}
The high computational cost of re-ranking documents with pre-trained transformers has recently received attention~\cite{hofstaetter:2019}. Even for pointwise approaches, the inference overhead can be prohibitive for practical applications~\cite{zhang:2021}. There are two ideas to improve the efficiency of neural re-rankers:
\Ni
improving the efficiency of the ranking model, and
\Nii
reducing the required number of inferences.

Approaches to the former include early-exiting from inference by intermediate between-layer classification in BERT-like models~\cite{xin:2020}, model distillation~\cite{hofstaetter:2020,gao:2020}, or improved dense representations~\cite{tang:2020}. For the latter, \citet{zhang:2021} propose to introduce filtering steps in multi-stage re-ranking pipelines. They utilize feature-based learning to rank to compute a set of candidate documents that is then re-ranked using a BERT-like neural model, increasing efficiency by a factor of up to~18 compared to an unfiltered baseline at the same effectiveness. But while document filtering has been studied for pointwise re-ranking, to our knowledge, filtering approaches for pairwise re-ranking have not been addressed to date.

\section{Sparsified Pairwise Re-Ranking}
\label{sec:methodological-approach}

In this section, we describe the steps we adapted in the mono-duo re-ranking pipeline (Figure~\ref{mono-duo-reranking-illustration}): sampling methods to select the to-be-compared document pairs (three methods, Section~\ref{method:sampling}), and aggregation methods that derive a ranking from the ranking preferences (five methods, Section~\ref{method:aggregation}). For completeness, we also briefly detail the steps adopted from the literature: initial retrieval, as well as pointwise and pairwise re-ranking (Section~\ref{method:initial-retrieval}).

\subsection{Sampling}\label{method:sampling}

Based on the top-$k$ results~$D_k$ of the pointwise re-ranking step of the mono-duo paradigm, we propose to sparsify the set~$C_{\mathit{all}}$ of all $k^2-k$~comparisons (no self-comparisons) and to use a sampled comparison set~$C \subset C_{\mathit{all}}$ as input for the pairwise re-ranking step. The goal is to minimize the size of~$C$ and thus the effort of pairwise re-ranking without compromising the quality of the final ranking.

\begin{figure}
\includegraphics[width=\columnwidth]{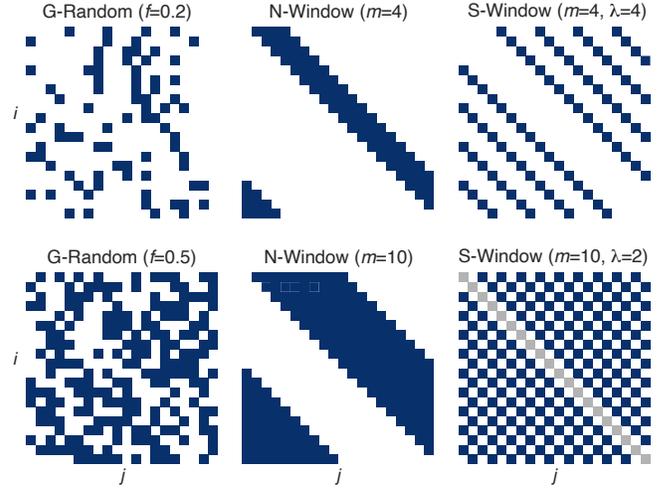}
\Description[Example comparison sets.]{Example comparison sets, where each examples is depicted as matrix over $D_k$ times $D_k$, where filled in cells indicate that a document pair has been sampled. Depicted are random and structured samples.}
\caption{Example comparison sets of different sampling procedures for 20~documents at different sampling rates. If comparison $(d_i, d_j)$ is sampled, cell $i,j$ is colored blue; the grey cell for S-Window illustrates the omitted case~$j=i$.} 
\label{fig:comparison-matrices-example}
\end{figure}

We distinguish random from structured sampling, with the main difference being their (non-)determinism. Independent random samples from a given~$C_{\mathit{all}}$ very likely contain different comparisons, but structured samples always choose the same comparisons. To be compatible with a variety of aggregation methods, a sampling must meet two requirements:
\Ni
each document is part of at least one comparison,
\Nii
each comparison is sampled at most once. Figure~\ref{fig:comparison-matrices-example} illustrates the three sampling methods introduced below for a document set~$D_k$ of size~$k=20$ at two sampling rates, one per line.

\paragraph{Global random sampling (G-Random).}
For each of the top-$k$ documents~$d\in D_k$ from the pointwise ranking, a fraction~$r\in(0...1]$ of the remaining $k-1$~documents is randomly selected for the comparison set~$C$ that then has the size~$|C|=\lfloor r\cdot (k^2-k)\rfloor$. 

\paragraph{Neighborhood window sampling (N-Window).}
A sliding window of size~$m \leq k-1$ is moved over the top-$k$ documents~$D_k$ from the pointwise ranking. For document~$d_i\in D_k$, its $m$~direct successors in the ranking are sampled for comparison. But since the last $m$~documents in the ranking of~$D_k$ have less than $m$~successors, we let the window ``wrap around'' to the top-ranked documents. For document~$d_i$, the $m$~sampled comparisons~$(d_i,d_j)\in C$ thus fulfill $j = 1 + (a~\mathrm{mod}~k)$ for $a\in\{i, \ldots, i+m-1\}$. The size of the sampled comparison set~$C$ is~$k \cdot m$ and each document is the first entry of $m$~comparisons and the second entry of another $m$~comparisons.

An assumption underlying the neighborhood sampling is that the ``global'' pointwise ranking is sensible but that ``local'' re-ranking leads to an improved effectiveness. If true, however, it would be plausible to stop the window for~$i > k-m$ rather than to wrap it around. However, pilot experiments have shown that this leads to poorer effectiveness, possibly because fewer comparisons are sampled at both the beginning and the end of a top-$k$ ranking.

\paragraph{Skip-window sampling (S-Window).}
N-Window samples from the ``local'' neighborhood in a ranking. To enable more ``global'' comparisons, we introduce a skip size~$\lambda \in \mathbb{N}^+$. For~$d_i\in D_k$, comparisons~$(d_i, d_j)$ to $m$~successors are sampled so that $j = 1 + (a~\mathrm{mod}~k)$ for $a \in \{ i + \lambda - 1, i + 2\lambda - 1, \ldots, i + m\lambda - 1 \}$; when $j = i$ for some~$a$, that comparison is not included in the sample. For~$\lambda=1$, S-Window corresponds to N-Window, and for~$\lambda=3$, for example, each document is compared to every third of its successors. The $\lambda$-skip deterministically controls the ``globality'' of a sample without increasing the amount~$|C|$ of sampled comparisons compared to N-Window.

\subsection{Aggregation}\label{method:aggregation}

For each comparison~$(d_i, d_j) \in C$, a pairwise model computes a preference probability~$p_{ij}$, which indicates how likely~$d_i$ should be ranked above~($p_{ij} \geq 0.5$) or below~$d_j$~($p_{ij} < 0.5$). From the probabilities computed for~$C$, an aggregation method derives a relevance value~$s_i$ for each document~$d_i$. We study five paradigmatically different aggregation methods.

\paragraph{KwikSort.}
As a baseline, we use the KwikSort method~\cite{ailon:2008}. It is an extension of the Quicksort algorithm for data with preferences and, in our case of top-$k$ ranking, has an expected number of comparisons in~$O(k\log k)$. First, a random document~$d_i$ is chosen to be the pivot. Then, all other documents are compared to the pivot placing the ones to be lower-ranked than~$d_i$ and the ones to be higher-ranked in separate subsets. These subsets are recursively ranked until a final ranking is obtained. KwikSort does not rely on a preceding sampling step; that the expected number of comparisons is in~$O(k\log k)$ is a feature of the dynamic aggregation itself.

\paragraph{Additive Aggregation.}
\citet{pradeep:2021} propose four different aggregation techniques based on preference probability summation. They find the symmetric sum of preference probabilities to yield the best effectiveness:
\begin{equation*}
s_i = \sum_{j \in 1 \ldots k}(p_{ij} + (1 - p_{ji}))\,.
\end{equation*}
However, in our samples, not all comparisons are present so that we replace missing summands~$p_{ij}$ or $(1 - p_{ji})$ by~$0$. 

\paragraph{Bradley-Terry Aggregation.}
The Bradley-Terry model~\cite{bradley:1952} infers a latent score~$s_i \in S$ for each document~$d_i \in D_k$ based on the preferences expressed in the sampled comparison set~$C$ using maximum-likelihood estimation. In its original form, exponential score functions were used, which corresponds to a logistic regression on pairwise data~\cite{agresti:2011} and can be expressed as:
\begin{equation*}
\mathcal{L}(S, C) = \sum_{d_i \succ d_j}\log\frac{e^{s_i}}{e^{s_i} + e^{s_j}} +
                    \sum_{d_i \prec d_j}\log\frac{e^{s_j}}{e^{s_i} + e^{s_j}}\,.
\end{equation*}
Here, $d_i \succ d_j$ denotes all comparisons $(d_i, d_j) \in C$ with $p_{ij} \geq 0.5$ and $d_i \prec d_j$ denotes all comparisons $(d_i, d_j) \in C$ with $p_{ij} < 0.5$. The unknown latent score set~$S$ is usually found via BFGS~optimization~\cite{fletcher:1987} so that a ranking according to the~$s_i$ violates as few of the preferences from the comparison sample~$C$ as possible.

\paragraph{Greedy Aggregation.}
\citet{cohen:1999} propose a greedy ordering algorithm that is proven to closely approximate the best total order in terms of the number of violated preferences. Algorithm~\ref{alg:greedy-sorting} shows its pseudocode. In every iteration, the document~$d_j$ with the highest ``potential''~$t_j$%
\footnote{The ``potential'' basically tallies~$d_i$'s ``wins'' against other documents compared to its ``losses'' in terms of preference probabilities.}
is appended to the re-ranking on the highest still unoccupied rank by setting score~$s_j$ accordingly. The potentials of the remaining documents are updated by canceling out the respective terms that include~$d_j$. With sampling, the comparison set is incomplete; missing probabilities~$p_{ij}$ are set to zero.

\begin{algorithm}[t]
\parbox[t]{4em}{\textbf{Input:}}Document set~$D_k$, preference probabilities~$p_{ij}$\\
\parbox[t]{4em}{\textbf{Output:}}Score~$s$ for each $d \in D_k$

\medskip
\lForEach{$d_i \in D_k$}{$t_i \gets \sum_{d_j \in D_k} p_{ij} - \sum_{d_j \in D_k} p_{ji}$}
\While{$D_k \neq \emptyset$}{
  $d_j \gets \argmax_{d_i \in D_k} t_i$\;
  $s_j \gets |D_k|$\;
  $D_k \gets D_k \setminus \{d_j\}$\;
  \lForEach{$d_i \in D_k$}{$t_i \gets t_i - p_{ij} + p_{ji}$}
}
\caption{Greedy aggregation of preferences~\cite{cohen:1999}.\label{alg:greedy-sorting}}
\end{algorithm}

\paragraph{PageRank Aggregation.}
A comparison set~$C$ induces a directed graph with~$D_k$ as nodes and comparisons as directed edges weighted with preference probabilities. We introduce a new aggregation method that computes the graph centrality measure PageRank~\cite{page:1999}, extended for weighted graphs~\cite{mihalcea:2004}, to rank the documents. The fundamental principle of PageRank is that nodes with incoming edges from nodes with high PageRank~scores should also receive high PageRank~scores. The respective PageRank-style aggregation of a score~$s_i$ for a document~$d_i$ then is
\begin{equation*}
s_i = \gamma \cdot \frac{1}{|D_k|} + (1-\gamma) \cdot\hspace{-1em}\sum_{(d_j, d_i) \in C} \frac{p_{ji}}{\sum_{l\in[1,k]}p_{jl}}\cdot s_j\,,
\end{equation*}
where using the components with the damping factor $\alpha\in[0,1]$ ensure convergence when computing the PageRank~scores iteratively.

\subsection{Initial Retrieval and Re-ranking}\label{method:initial-retrieval}

Following the experimental setup of \citet{pradeep:2021} closely (see Figure~\ref{mono-duo-reranking-illustration}), for each query, we first obtain an initial ranking using~BM25 (PyTerrier implementation~\cite{macdonald:2020}, default configuration). The top-1000 BM25~results are then re-ranked using monoT5~\cite{nogueira:2020} in the pointwise re-ranking step. For the top-50 monoT5~results, duoT5~\cite{pradeep:2021} infers preference probabilities in the second step of pairwise re-ranking, after sampling. For both, monoT5 and duoT5, we apply the largest available pre-trained version.%
\footnote{monoT5: \url{https://huggingface.co/castorini/monot5-3b-msmarco} \\\hspace*{0.2em} duoT5: \url{https://huggingface.co/castorini/duot5-3b-msmarco}}
We use~T5 instead of BERT~variants, as T5~has been shown to be more effective~\cite{lin:2021}. To avoid repeated inferences in our experiments, all $k^2-k$~pairwise preference probabilities are cached once for each query.

The maximum input length of transformer models is limited, so that a representative passage has to be chosen from each document for inference. Following the method of \citet{dai:2020}, we split each document into fixed-length non-overlapping passages of about 250~words (using the TREC~CAsT~Y4 tools;%
\footnote{\url{https://github.com/grill-lab/trec-cast-tools}}
splits at sentence boundaries). Fixed-length passages have been shown to be more effective than variable-length passages~\cite{kaszkiel:1997}. In our pilot experiments, using the first passage was the most effective heuristic, so that we use them for preference probability inference.

\section{Experimental Setup}\label{sec:setup}

In this section, we introduce our evaluation measures, detailing in particular measures for consistency, complementarity, and transitivity of the aggregated relevance scores, and recap the used datasets. 

\begin{figure*}[ht!]
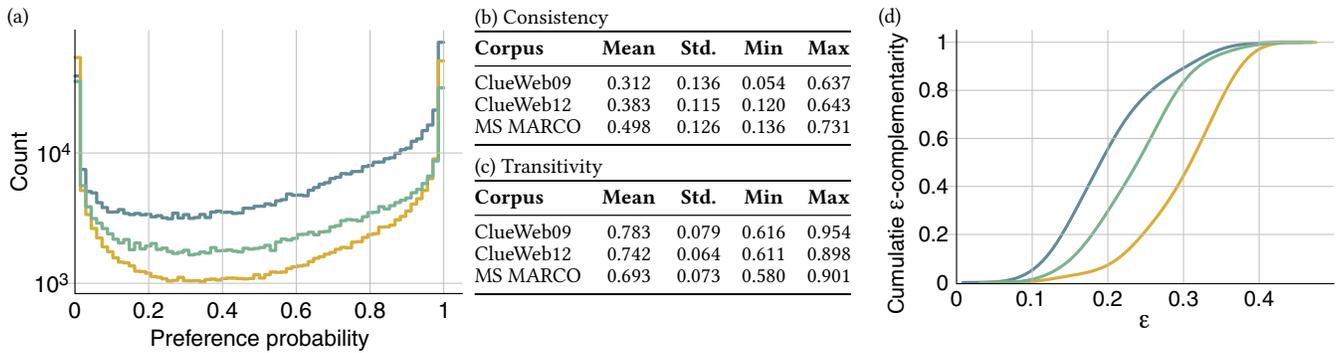

\parbox[t]{0.34\textwidth}{
\small%
(a)\\
\includegraphics[width=.34\textwidth]{plot-preference-probability-distribution-per-corpus}%
\Description[Distribution of the pairwise duoT5~preference probabilities on the ClueWeb09, ClueWeb12, and the passage dataset of MS~MARCO.]{Distribution of the pairwise duoT5~preference probabilities on the ClueWeb09, ClueWeb12, and the passage dataset of MS~MARCO.}}
\hspace*{0.25em}%
\parbox[t]{0.28\textwidth}{%
\small%
\vskip-1.6ex
\renewcommand{\tabcolsep}{4pt}
\renewcommand{\arraystretch}{0.84}
\begin{tabular}{@{}lcccc@{}}
  \multicolumn{5}{@{}l@{}}{(b) Consistency}                                          \\
\toprule
  \bfseries Corpus & \bfseries Mean & \bfseries Std. & \bfseries Min & \bfseries Max \\
\midrule
  ClueWeb09        &     0.312      &     0.136      &     0.054     &     0.637     \\
  ClueWeb12        &     0.383      &     0.115      &     0.120     &     0.643     \\
  MS~MARCO         &     0.498      &     0.126      &     0.136     &     0.731     \\
\bottomrule
\end{tabular}%
\vskip1ex
\begin{tabular}{@{}lcccc@{}}
  \multicolumn{5}{@{}l@{}}{(c) Transitivity}                                         \\
\toprule
  \bfseries Corpus & \bfseries Mean & \bfseries Std. & \bfseries Min & \bfseries Max \\
\midrule
  ClueWeb09        &     0.783      &     0.079      &     0.616     &     0.954     \\
  ClueWeb12        &     0.742      &     0.064      &     0.611     &     0.898     \\
  MS~MARCO         &     0.693      &     0.073      &     0.580     &     0.901     \\
\bottomrule
\end{tabular}%
}
\hspace*{1em}%
\parbox[t]{0.34\textwidth}{%
\small%
(d)\\
\includegraphics[width=.34\textwidth]{plot-epsilon-complimentarity-per-corpus}
\Description[Cumulative complementarity of preference probabilities.]{Cumulative complementarity of preference probabilities.}}
\caption{%
(a)~Distribution of the pairwise duoT5~preference probabilities for the comparison set~$C_{\mathit{all}}$ of the top-50 pointwise results per corpus (log-scaled y-axis).
(b)~Preference probability consistency,
(c)~transitivity, and
(d)~cumulative $\varepsilon$-complementarity over~$\varepsilon$. In the plots, the corpora are color-coded as ClueWeb09 {\color{c_blue}\rule{15pt}{6.75pt}}, ClueWeb12 {\color{c_green}\rule{15pt}{6.75pt}}, and MS~MARCO (Passage) {\color{c_yellow}\rule{15pt}{6.75pt}}.}
\label{fig:dataset-characteristics}
\end{figure*}

\subsection{Evaluation Measures}

We follow similar studies of the mono/duoT5~models~\cite{pradeep:2021} and use nDCG@10~\cite{jarvelin:2002} to evaluate the retrieval effectiveness. When re-ranking the top-$k$ results of a pointwise model, the $k^2 - k$~comparisons~$C_{\mathit{all}}$ usually performed by a pairwise model may result in inconsistent preference probabilities 
\Ni
at the level of a document pair and
\Nii
at the level of document triples.
We further examine these potential inconsistencies, as they can ``complicate'' the aggregation step and affect the retrieval effectiveness. At the document pair level, one would expect~$p_{ij} \approx 1 - p_{ji}$ but pairwise models do not guarantees this and may predict both~$d_i \succ d_j$ for the input pair $(d_i, d_j)$ and $d_j \succ d_i$ for the input pair $(d_j, d_i)$, or vice versa, where $d_i \succ d_j$ denotes a ranking preference of the left document~$d_i$ over the right one~$d_j$. At the document triple level, transitivity may be violated as a model may predict~$d_i \succ d_j$ and~$d_j \succ d_l$ but~$d_l \succ d_i$.

The consistency of an all-$(k^2 - k)$-pairs comparison set~$C_{\mathit{all}}$ at the level of document pairs is the fraction of pairs~$(d_i, d_j) \in C_{\mathit{all}}$ for which~$p_{ij} \geq 0.5$ and~$p_{ji} < 0.5$:
\begin{equation*}
\mathit{consistency}(C_{\mathit{all}}) \ = \ \frac{|\{(d_i, d_j) \in C_{\mathit{all}} : p_{ij} \geq 0.5 \text{ and } p_{ji} < 0.5\}|}{|C_{\mathit{all}}|}.
\end{equation*}
While consistency captures the comparison direction, also the numerical complementarity of how close $p_{ij} + p_{ji}$ is to the ``ideal''~$1$ can be interesting. Thus, we also measure the \mbox{$\varepsilon$-complementarity} with respect to a margin of error~$\varepsilon$ as: 
\begin{equation*}
\varepsilon\text{-}\mathit{complementarity}(C_{\mathit{all}}) \ = \ \frac{|\{(d_i, d_j) \in C_{\mathit{all}} : |p_{ij} + p_{ji} - 1| < \varepsilon \}|}{|C_{\mathit{all}}|}.
\end{equation*}
Finally, the transitivity of~$C_{\mathit{all}}$ measures the fraction of document triples for which the pairwise comparisons are transitive:
\begin{equation*}
\mathit{transitivity}(C_{\mathit{all}}) \ = \ \frac{|T|}{|T| + |I|}\,, \text{ where}
\end{equation*}
$$
\renewcommand{\arraycolsep}{2pt}
\begin{array}{rcl}
 T & = & \{(d_i, d_j, d_l) : p_{ij} \geq 0.5, \ p_{jl} \geq 0.5, \text{ and } \ p_{il} \geq 0.5\} \;\cup\\
   &   & \{(d_i, d_j, d_l) : p_{ij} < 0.5, \ p_{jl} < 0.5, \text{ and } \ p_{il} < 0.5\} \qquad\text{and}\\[1ex]
 I & = & \{(d_i, d_j, d_l) : p_{ij} \geq 0.5, \ p_{jl} \geq 0.5, \text{ but } \ p_{il} < 0.5\} \;\cup\\
   &   & \{(d_i, d_j, d_l) : p_{ij} < 0.5, \ p_{jl} < 0.5, \text{ but } \ p_{il} \geq 0.5\}\,.
\end{array}
$$

The more the $\varepsilon$-complementarity for some small~$\varepsilon$ and the more the transitivity of some $C_{\mathit{all}}$ approach~1, the more a total order between the documents is implied that probably can also be derived from some smaller comparison sample~$C \subset C_{\mathit{all}}$.

\subsection{Evaluation Data}

We employ three standard retrieval corpora in our experiments: the ClueWeb09, the ClueWeb12, and the MS~MARCO passage corpus.

\paragraph{ClueWeb09.}
The ClueWeb09 corpus%
\footnote{\url{http://lemurproject.org/clueweb09.php/}}
consists of 1~billion documents crawled between January and February~2009. It was used for the ad-hoc search tasks of the TREC Web tracks~2009--2012~\mbox{\cite{clarke:2009,clarke:2010,clarke:2011,clarke:2012}}, where 70,575~graded relevance judgments were \mbox{collected} on~a \mbox{4-point} scale for 200~topics (avg.~356~judgments per topic).

\paragraph{ClueWeb12.}
The ClueWeb12 corpus%
\footnote{\url{http://lemurproject.org/clueweb12.php/}}
consists of 733~million documents crawled between April and May~2012. It was used for the ad-hoc search tasks of the TREC Web tracks~2013/14~\mbox{\cite{collins:2013,collins:2014}}, where 28,116~graded relevance judgments were collected on a \mbox{4-point} scale for 100~topics (avg.~281~judgments per topic).

\paragraph{MS~MARCO (Passage).}
The MS~MARCO passage corpus~\cite{nguyen:2016} consists of 8.8~million passages extracted from Bing search engine results. It was used for the passage ranking task of the TREC Deep Learning tracks~2019/20~\cite{craswell:2020,craswell:2021}, where 20,646 graded relevance judgments were collected on a 4-point scale for 97~topics (avg.~213~judgments per topic). With this corpus, we replicate the experimental setup of \citet{pradeep:2021}.

\paragraph{Remark}
In our evaluation of re-ranking results, we only consider judged documents. Evaluation scores calculated excluding unjudged documents correlate well with evaluations including them~\cite{sakai:2007}.
\section{Evaluation Results}\label{sec:results}

We conduct two experiments to evaluate the suitability of sampling and aggregation methods for efficient pairwise re-ranking. The first experiment (Section~\ref{sec:eval-comparison-properties}) explores the properties of the comparison sets inferred for each of the three corpora. This supplies context to the ranking effectiveness evaluation in the second experiment (Section~\ref{sec:eval-sampling-aggregation}), in which we analyze rankings for different combinations of samplers and aggregators.

\begin{figure*}[ht]
\includegraphics[width=\textwidth]{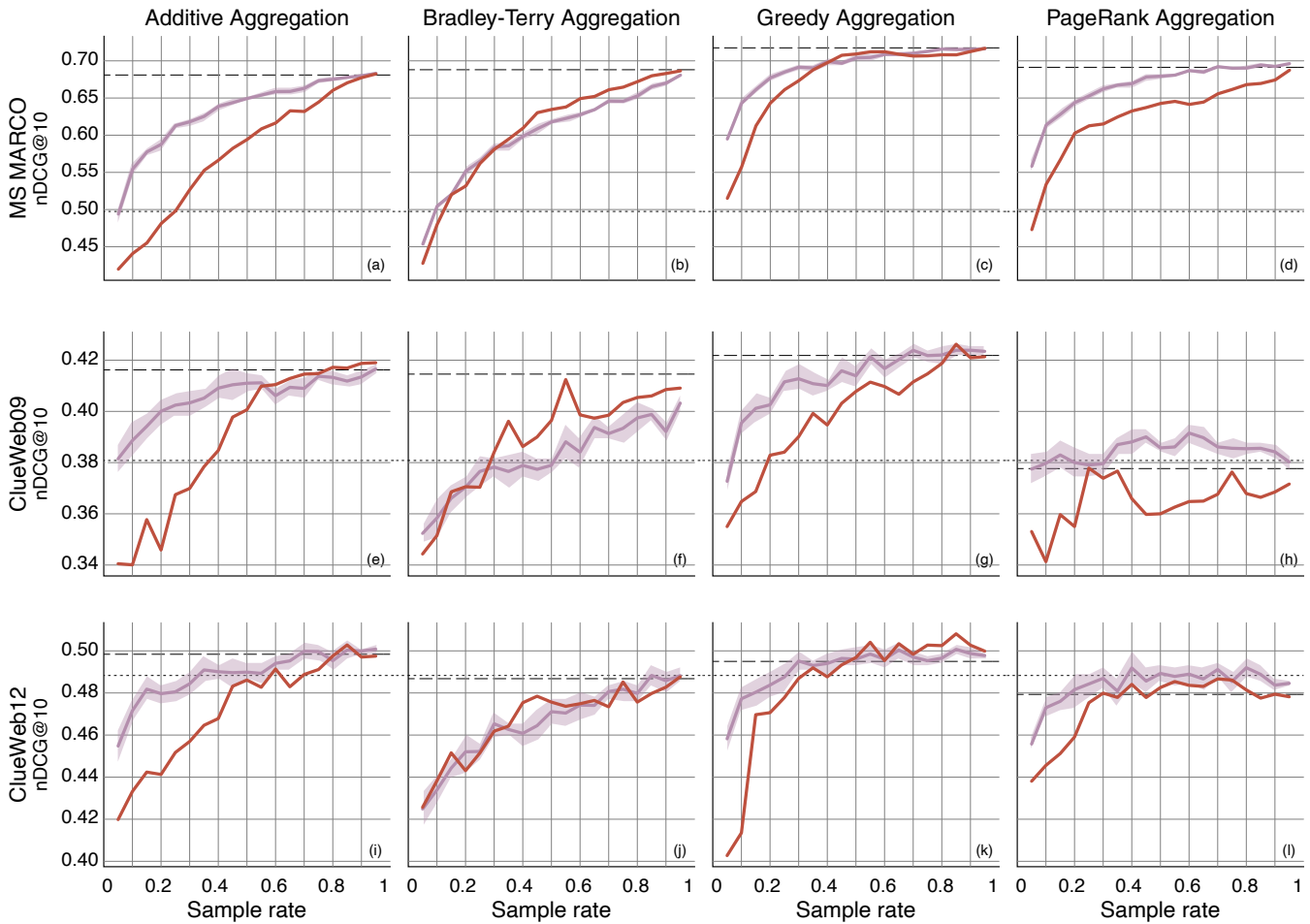}
\Description[Effectiveness measured as nDCG@10 for each aggregator at different sampling rates for the G-Random and N-Window sampling strategies.]{Effectiveness measured as nDCG@10 for each aggregator at different sampling rates for the two sampling strategies.}
\caption{Effectiveness measured as nDCG@10 on three corpora (ClueWeb09, ClueWeb12, MS~MARCO) for each aggregator and the two samplers G-Random~{\color{c_purple}\rule{15pt}{6.75pt}} and N-Window~{\color{c_red}\rule{15pt}{6.75pt}} with different sampling rates. Dotted: pointwise re-ranking, dashed: unsampled~$C_{\mathit{all}}$.}
\label{fig:effectiveness}
\end{figure*}

\subsection{Evaluation of Pairwise Prediction Properties}
\label{sec:eval-comparison-properties}

For each topic from each corpus, we derive the duoT5~preference probabilities for the set~$C_{\mathit{all}}$ of all $k^2 - k$~pairwise comparisons for the top-50 results of the pointwise re-ranking and compute the statistics and measures per corpus shown in Figure~\ref{fig:dataset-characteristics}.

The preference probabilities are highly skewed towards the extremes of the scale (cf.\ Figure~\ref{fig:dataset-characteristics}a): for the majority of document pairs, the preference probability is approximately zero or one. This effect is stronger for the MS~MARCO passage corpus (on which the model was trained) than for the ClueWeb corpora. Interestingly, the score distributions are not symmetric, but are slightly skewed towards~1.0 for all three corpora. Since the comparison set~$C_{\mathit{all}}$ contains both comparison directions for every pair, the observed skew directly suggests to further inspect how consistent, transitive, and complementary the preferences are for document pairs or triples.

Indeed, on average, only between half (MS~MARCO) and a third of the comparisons (ClueWeb09) are consistent in their direction (cf.\ Figure~\ref{fig:dataset-characteristics}b). Some variation across topics exists, yet the consistency is rather low in the majority of the topic-wise comparison sets. Further, also the cumulative $\epsilon$-complementarity (cf.\ Figure~\ref{fig:dataset-characteristics}d) confirms that the preferences of the pairwise duoT5~model are not that complementary (i.e., $p_{ij} + p_{ji} \neq 1$) for a document pair's two possible input orders. Only for a rather large value of~$\varepsilon = 0.4$ all probabilities for pairs in all corpora are $\varepsilon$-complementary (i.e., $|p_{ij} + p_{ji} - 1| \neq 0.4$), and more than half of the pairs require an $\epsilon$-value between~0.3 (MS~MARCO) and~0.2~(ClueWeb09). For all corpora, almost no comparison pairs reach a complementarity of~$\epsilon < 0.1$. Also the transitivity rates are consistently between~0.7 and~0.8 across all corpora with very little variation per topic. These observations (consistency and transitivity not that high) suggest that the comparison-based KwikSort aggregation will not output the most effective re-ranking and that very likely more than $O(k \log k)$ comparison pairs are needed in a sample for the other aggregation methods to ``work around'' the low consistency and lacking complementarity using more information.

\enlargethispage{-\baselineskip}
\subsection{Evaluation of Ranking Effectiveness}
\label{sec:eval-sampling-aggregation}

To evaluate the effectiveness of different combinations of sampling and aggregation methods on all three corpora, we simulate runs on sparsified comparison sets at sample rates ranging from~0.05 to~0.95 in steps of~0.05. Each simulation is repeated ten times and the effectiveness is averaged to account for random variation in both the sampling and the aggregation step. In addition, baseline runs for each aggregator use the full comparison sets~$C_{\mathit{all}}$ per topic without any sampling. In total, 4950~runs are simulated. The pointwise ranking to be re-ranked by a pairwise model achieves nDCG@10~scores of 0.38~(ClueWeb09), 0.49~(ClueWeb12) and~0.50~(MS~MARCO).

\paragraph{Effectiveness of KwikSort.} 
From the observations in Section~\ref{sec:eval-comparison-properties} (low consistency and low transitivity), it is clear that KwikSort with its pivot-based dynamic sampling of a rather ``few'' $O(k \log k)$ comparisons will not be able to achieve a really good effectiveness. Indeed, the nDCG@10~scores of KwikSort of 0.34~(ClueWeb09), 0.39~(ClueWeb12), and 0.42 (MS~MARCO) are even lower than the pointwise effectiveness. We tested different pivot selection methods but all resulted in a similarly bad effectiveness. Using KwikSort for pairwise rank aggregation can thus not be recommended in settings with inconsistent and intransitive comparisons.

\begin{figure*}[t]
\includegraphics[width=\textwidth]{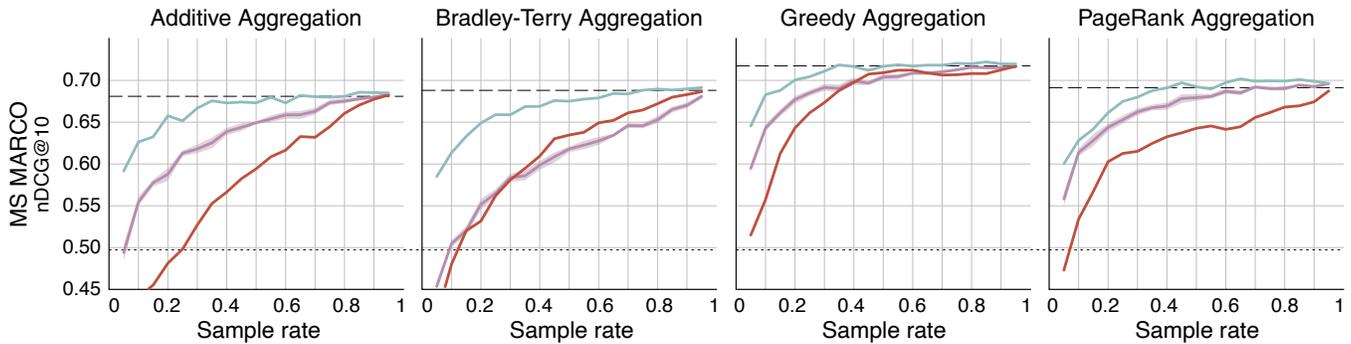}
\Description[nDCG@10 on MS~MARCO for each aggregator at different sampling rates for three sampling strategies.]{nDCG@10 on MS~MARCO for each aggregator at different sampling rates for three sampling strategies.}
\caption{nDCG@10 on MS~MARCO for each aggregator at different sampling rates for S-Window {\color{c_blue}\rule{15pt}{6.75pt}}, G-Random {\color{c_purple}\rule{15pt}{6.75pt}}, and N-Window {\color{c_red}\rule{15pt}{6.75pt}}. Dotted line: pointwise ranking, dashed line: effectiveness on unsampled~$C_{\mathit{all}}$.}
\label{fig:gridsearch}
\end{figure*}

\paragraph{Effectiveness on the full comparison set~$C_{\mathit{all}}$.}
Figure~\ref{fig:effectiveness} shows the nDCG@10 of the non-KwikSort aggregation methods on each corpus. The dotted and dashed horizontal lines indicate the baseline effectiveness of the pointwise ranking and the pairwise re-ranking, respectively, when using the respective aggregation method on the complete comparison set~$C_{\mathit{all}}$ without sampling. Since the pointwise and the $C_{\mathit{all}}$-aggregation effectiveness are thus independent of the sampling rate, they are depicted as horizontal lines.

Using the full comparison set~$C_{\mathit{all}}$ without any sampling, greedy aggregation yields the most effective re-rankings for the ClueWeb09 (Subplot~\ref{fig:effectiveness}c) and the MS~MARCO passage corpus (Subplot~\ref{fig:effectiveness}g) while additive aggregation is the most effective on the ClueWeb12 (Subplot~\ref{fig:effectiveness}i). On MS~MARCO (first row of Figure~\ref{fig:effectiveness}), all aggregation methods are approximately equally effective when using the full comparison set~$C_{\mathit{all}}$ (dashed lines show about the same~nDCG@10). On the ClueWeb corpora, though, PageRank aggregation is less effective than the pointwise ranking (Subplots~\ref{fig:effectiveness}h and~l; dashed line below dotted line) and also Bradley-Terry aggregation struggles on the ClueWeb12 (Subplot~\ref{fig:effectiveness}j). Only additive and greedy aggregation always improve upon the pointwise ranking when using the full comparison set~$C_{\mathit{all}}$ (Subplots~\ref{fig:effectiveness}e,~g,~i, and~k) but the improvement is smaller on the ClueWeb corpora. That the effectiveness and the improvement over the pointwise ranking are the highest on the MS~MARCO passage corpus is not surprising since the duoT5~re-ranking model was trained on MS~MARCO, and since the TREC Web track relevance judgments used to evaluate the effectiveness on the ClueWeb corpora are at the document level, while we only rank one passage per document due to input length limitations.

\paragraph{Effectiveness with G-Random and N-Window sampling.} 
The color-coded curves in the plots of Figure~\ref{fig:effectiveness} show the effectiveness of the different aggregation methods with G-Random or N-Window sampling at different sampling rates from the full comparison set~$C_{\mathit{all}}$. Four trends are apparent across all corpora.

First, N-Window sampling results in less effective re-rankings than G-Random sampling in nearly all cases, especially at smaller sampling rates. This effect is particularly noticeable for additive aggregation (Subplots~\ref{fig:effectiveness}a,~e, and~i). One reason probably is that the more ``local'' comparisons of N-Window are likely to yield less extreme comparison probability differences that decrease the overall separability of document pairs in sparse sampling setups. Also, inconsistencies in pairwise judgments are more likely for ``local'' pairs. Overall, this indicates that the global context of documents (which is better represented by G-Random) is important to obtain effective re-rankings via aggregation.

Second, greedy aggregation is the most effective aggregation method for both, G-Random and N-Window sampling (comparing Subplots~\ref{fig:effectiveness}c,~g, and~k to the rest). It is also the only aggregation method for which the effectiveness on the full~$C_{\mathit{all}}$ is reached by some sparsified comparison sets.

Third, the effectiveness degradation is not linear with respect to the sample rate (all subplots), but drops sharply below~15--20\%. This suggests a lower bound of comparisons needed to derive good rankings from the pairwise comparisons of~duoT5 which lack in consistency and transitivity.

Fourth, Bradley-Terry- and PageRank-aggregated re-rankings often are the least effective (Subplots~\ref{fig:effectiveness}b,~f, and~j, as well as~d,~h, and~l). A possible reason for Bradley-Terry is similar to the bad effectiveness of KwikSort-aggregated re-rankings: Bradley-Terry only takes the direction of a comparison into account but not the magnitude of the respective probability. With the inconsistent probabilities of~duoT5 that lead to inconsistent comparison directions, Bradley-Terry cannot derive good final re-rankings---just like KwikSort. In case of PageRank aggregation, also the inconsistent probabilities that are used as edge weights, might ``confuse'' the actual derivation of the PageRank~scores.

\paragraph{Effectiveness with S-Window sampling.} 
To find a good value for~$\lambda$ in S-Window sampling, we run a grid search over $\lambda = 2\ldots 15$, separately for all sample rates (0.05 to~0.95 in steps of~0.05). We use five-fold cross validation to determine the best choice on the MS~MARCO corpus, as the overall effectiveness gains on the ClueWeb corpora were too small to meaningfully distinguish between setups. Figure~\ref{fig:gridsearch} shows the nDCG@10 effectiveness on MS~MARCO of the run with the optimal $\lambda$-value for each sample rate. The best runs for G-Random and N-Window are also shown for reference.

Re-rankings aggregated from S-Window samples are more effective by a margin for each of the aggregation methods at all sampling rates. The combination of S-Window sampling with greedy aggregation allows for a rather stable effectiveness down to sampling only~$30\%$ of the comparisons. Even when using an order of magnitude fewer comparisons~(i.e., $\approx 10\%$ of~$C_{\mathit{all}}$), a competitive effectiveness is achieved (nDCG@10 only~0.04 less).

The best values for~$\lambda$ are between~7 and~10 in most cases; they are not correlated with the window size~(Pearson's~$\bar{\rho}=0.04$). Already the better effectiveness of aggregated rankings using G-Random sampling over N-Window sampling suggests that the global context is important when sampling comparisons. Also the rather large best-working $\lambda$-values for S-Window sampling corroborate this since even for small sample sizes~$m$ they ensure that the sampled comparisons cover a pretty ``global'' context. For large sample sizes~$m$, $\lambda$~is not as important as the sample then already covers a larger amount of the full comparison set~$C_{\mathit{all}}$.

\paragraph{Minimal sampling rates.}
Table~\ref{tab:significance} shows the minimum attainable sampling rates on the MS~MARCO corpus for which each combination of sampling and aggregation method is not significantly less effective in terms of~nDCG@10 than the respective aggregation on the full comparison set~$C_{\mathit{all}}$. Per aggregator, the difference of the runs for each of the 19~sampling rates is tested against the run that aggregates a ranking from the full comparison set~$C_{\mathit{all}}$ using a paired Student's t-test with an $\alpha$-level of~0.05 and Bonferroni correction for the multiple tests. For G-Random with potentially different effectiveness scores for the 10~runs per sampling rate, we use the least effective run per sampling rate in terms of~nDCG@10 to increase the overall confidence in case of observed differences.

Among the sampling strategies, S-Window achieves the by far lowest sampling rates per aggregator without hurting the retrieval effectiveness too much. About the same effectiveness is possible with S-Window for additive, greedy, and PageRank aggregation with just one third of the usually used unsampled comparisons. G-Random and N-Window need more comparisons with any aggregation method to achieve the same re-ranking effectiveness and, in fact, only lead to some substantial savings compared to the unsampled~$C_{\mathit{all}}$ for PageRank aggregation (G-Random) or greedy aggregation (N-Window).

Among the aggregation strategies, additive and Bradley-Terry aggregation are the least effective and all sampling methods need larger sample rates for Bradley-Terry than for the other aggregators. Greedy aggregation leads to the best effectiveness and G-Random and N-Window achieve their lowest sampling rates without effectiveness loss for greedy aggregation.

Overall, the best combination in terms of effectiveness and sampling rate is greedy aggregation with S-Window sampling: with about one third of the usual $k^2 - k$ comparisons, the best effectiveness can be reached.

\begin{table}[t]
\centering
\small
\renewcommand{\tabcolsep}{4.5pt}
\caption{Effectiveness on the MS~MARCO corpus as nDCG@10 for the full comparison set~$C_{\mathit{all}}$ and the lowest similarly effective sampling rate (non-significant nDCG@10~difference; delta in brackets) per sampling method and aggregator. Bonferroni-correction for all (incl.\ hidden) tests per row.}
\begin{tabular}{@{}lcccc@{}}
\toprule
  \bfseries Aggregator &     \bfseries nDCG@10\phantom{0}      &                                \multicolumn{3}{c@{}}{\bfseries Lowest Similarly Effect.\ Sampl.\ Rate}                                \\
  \cmidrule(l){3-5}    & {\color{gray}\footnotesize Unsampled $C_{\mathit{all}}$} &                 S-Window                 &                 G-Random                 &                 N-Window                 \\
\midrule
  Additive             &                 0.691                 & 0.35 {\color{gray}\footnotesize(-0.014)} & 0.85 {\color{gray}\footnotesize(-0.019)} & 0.95 {\color{gray}\footnotesize(-0.004)} \\
  Bradley-Terry        &                 0.691                 & 0.50 {\color{gray}\footnotesize(-0.012)} & 1.00 {\color{gray}\footnotesize(-0.000)} & 0.90 {\color{gray}\footnotesize(-0.008)} \\
  Greedy               &                 0.707                 & 0.30 {\color{gray}\footnotesize(-0.013)} & 0.85 {\color{gray}\footnotesize(-0.006)} & 0.50 {\color{gray}\footnotesize(-0.010)} \\
  PageRank             &                 0.695                 & 0.30 {\color{gray}\footnotesize(-0.016)} & 0.65 {\color{gray}\footnotesize(-0.012)} & 0.95 {\color{gray}\footnotesize(-0.004)} \\
\bottomrule
\end{tabular}
\label{tab:significance}
\end{table}

\section{Conclusion}
\label{sec:conclusion}

In this paper, we analyze several methods to substantially reduce the quadratic number of document comparisons usually conducted in pairwise re-ranking with transformers. To this end, we introduce a sampling step at the beginning of the pairwise re-ranking and adapt the aggregation step to derive relevance scores for a re-ranking from smaller samples of the pairwise comparisons. By comparing combinations of three sampling methods and five aggregation methods, we show that only one third of the comparisons are needed to achieve competitive re-ranking effectiveness and that also an order of magnitude less comparisons still can yield only a very slightly decreased effectiveness. 

Compared to the usually applied additive rank aggregation without sampling, our new combination of skip-window sampling with greedy aggregation achieves an even better effectiveness at only about one third of the comparisons. When tolerating a very slight loss in effectiveness, even an order of magnitude fewer comparisons suffice. The more local exhaustive window sampling method leads to less effective rankings for which larger samples are needed than for skip-window or a global random sample. This suggests that a good sample of pairwise comparisons should not just sample from a very local environment per rank in the pointwise ranking.

Sparsification in pairwise re-ranking opens up new areas of research. Of the sampling paradigms evaluated (random vs.\ structured and local vs.\ global), the global structured sampling works better than the random one. Still, the samples are static in the sense that the sample is pre-computed before aggregation. Dynamic sampling techniques, in which new comparisons could be selected even during aggregation could merit further analyses. Sparsification could also be used to increase the depth of the pairwise re-ranking rather than its efficiency. Instead of minimizing the comparison budget for a fixed depth~$k$, a fixed comparison budget can be used to maximize~$k$. For example, the usually recommended depth~$k=50$ requires 2,450~comparisons ($k^2-k$) for traditional pairwise re-ranking. A sampling rate of~30\%~(or 10\%) now allows a re-ranking depth of~$k=90$~($k=157$) for a budget of about 2,450~comparisons. This may have a strong effect for recall-intensive retrieval tasks.

\balance
\bibliographystyle{ACM-Reference-Format}
\begin{raggedright}

\end{raggedright}

\end{document}